\documentclass[a4paper,showpacs,prb,twocolumn,superscriptaddress]{revtex4}

\usepackage{amsfonts}
\usepackage{graphicx}
\usepackage{color}
\usepackage{amsmath}
\usepackage{amssymb}
\usepackage{latexsym}
\usepackage{psfrag}

\begin{document}
\title{Electron-electron interactions in antidot-based Aharonov-Bohm interferometers}
\author{S. Ihnatsenka}
\affiliation{Department of Physics, Simon Fraser University, Burnaby, British Columbia, Canada V5A 1S6}
\author{I. V. Zozoulenko}
\affiliation{Solid State Electronics, Department of Science and Technology (ITN), Link\"{o}ping University, 60174 Norrk\"{o}ping, Sweden}
\author{G. Kirczenow}
\affiliation{Department of Physics, Simon Fraser University, Burnaby, British Columbia, Canada V5A 1S6}

\begin{abstract}
We present a microscopic picture of quantum transport in quantum antidots in the quantum Hall regime taking electron interactions into account. We discuss the edge state structure, energy level evolution, charge quantization and linear-response conductance as the magnetic field or gate voltage is varied. Particular attention is given to the conductance oscillations due to Aharonov-Bohm interference and their unexpected periodicity. To explain the latter we propose the mechanisms of scattering by point defects and Coulomb blockade tunneling. They are supported by self-consistent calculations in the Hartree approximation, which indicate pinning and correlation of the single-particle states at the Fermi energy as well as charge oscillation when antidot-bound states depopulate. We have also found interesting phenomena of anti-resonance reflection of the Fano type.
\end{abstract}

\maketitle

\section{Introduction}

An antidot is a potential energy hill in a two-dimensional electron gas (2DEG) formed in a GaAs/AlGaAs heterostructure by a negative voltage applied on a surface gate.\cite{Sim_review} It is often regarded as an artificial repulsive impurity and thus considered to be the inverse of a quantum dot. By applying a magnetic field perpendicular to the 2DEG, antidots have been extensively and intensively studied to understand
edge state transport in the quantum Hall regime,\cite{Jain88, Hwang91, Ford94, Sachrajda94, George94, Kataoka00, antidot}
charging in open systems and its influence on Aharonov-Bohm (AB) interference,\cite{Ford94, Kataoka00, Sim03, Kataoka99, Kataoka03}
fractionally quantized charge of Laughlin quasiparticles,\cite{Goldman95} 
nonabelian statistics of the fractional quantum Hall $5/2$ state\cite{DasSarma05}
 and others.\cite{Sim_review}  
Though electron transport in antidots seemed to be well understood, recent experiments of Goldman \textit{et al.} have revealed new unexpected features such as multiple periodicity of the Aharonov-Bohm conductance oscillations.\cite{Goldman08}

In a magnetic field, quantum Hall edge channels form closed pathways encircling an antidot.\cite{Sim_review, Davies_book} They are separated from extended edge channels propagating along device boundaries by quantum point contact (QPC) constrictions; see the insets in Fig. \ref{fig:GvsB}. At a given magnetic field there are $f_{leads}$ propagating edge states in the leads at the Fermi energy. The electron density in the constrictions is smaller than in the leads, and hence only the lowest $f_{c}$ states are fully transmitted, whereas the remaining highest $f_{leads}-f_{c}$ states are partially or fully reflected. A typical conductance of the AB interferometer as a function of magnetic field exhibits a step-like structure with plateaus separated by wide transitions regions.\cite{Hwang91, Ford94, Sachrajda94, George94, Goldman08} At very low magnetic fields the conductances of the QPC constrictions are additive and behave like classical Ohmic resistors.\cite{George90} When the magnetic field increases the conductance evolves from classical to quantum behavior with the single-particle levels condensed into degenerated Landau levels (LLs). The step-like dependence of the conductance reflects successive depopulation of the LLs in the constrictions.\cite{Davies_book} The plateau regions correspond to the field range where the QPC openings are fully transparent (the transmission coefficient through an individual QPC is integer, $T\cong f_{c}$), and transition regions between these plateaus correspond to the partially transparent QPC openings (the transmission coefficient is non-integer, $f_{c}<T<f_{c}+1)$. When a single-particle state of the antidot-bound edge channel coincides with the Fermi energy $E_F$, it provides a pathway for scattering from an edge channel on one side of the sample to an edge channel on the opposite side. Thus, it gives rise to pronounced AB conductance oscillations in the transition regions between the plateaus.


According to the conventional theory of the Aharonov-Bohm interferometer\cite{Davies_book} its conductance shows a peak each time the enclosed flux $\phi = BS$ changes by the flux quantum $\phi_{0} = h/e$. Thus, the conductance of the interferometer as a function of the magnetic field exhibits the periodicity
\begin{equation}
	\Delta B=\frac{\phi _{0}}{S}.  
	\label{period_1}
\end{equation}%
Here $S=\pi r^2$ is the area enclosed by a circular antidot-bound state of radii $r$, which approximates the geometrical area the antidot gate, Fig. \ref{fig:GvsB}. The enclosed flux through the interferometer can also be varied at a fixed magnetic field by changing an antidot gate voltage $V_{adot}$. In the case when the area changes linearly with the change of the gate voltage, $\Delta S=\alpha \Delta V_{adot},$ the expected periodicity is
\begin{equation}
	\Delta V_{adot}=\frac{\phi _{0}}{\alpha B}.  
	\label{period_Vg}
\end{equation}

Since the experimental study of Hwang \textit{et al.}\cite{Hwang91} the interpretation of Aharonov-Bohm oscillations based on Eq. \eqref{period_1} has been widely accepted.\cite{Sim_review} Measuring the period $\Delta B$, the radii of the edge states circulating around the antidot can be deduced. This gives valuable information about actual depletion region in two-dimensional electron gas (2DEG). However, in the recent experimental work of Goldman \textit{et al.}\cite{Goldman08}, it was reported that the periodicity of the AB oscillations as a function of the magnetic field depends on the number of fully transmitted states in the constriction $f_c$ and is then well described by the dependence
\begin{equation}
	\Delta B=\frac{1}{f_{c}}\frac{\phi _{0}}{S}, 
	\label{period_f_c}
\end{equation}%
which differs by a factor $1/f_c$ from the conventional formula \eqref{period_1}. On the other hand, the back-gate charge period $\Delta V_{adot}$ was found to be the same for all $f_c$, independent of the magnetic field in stark contrast to Eq. \eqref{period_Vg}.\cite{Goldman08} This departure from the conventional periodicity of the AB oscillations cannot be explained in a one-electron picture of non-interacting electrons. To account for it, it is necessary to consider electron interactions and/or Coulomb blockade (CB) charging effects. Coulomb interactions define a potentially important energy scale because even rough estimation gives Coulomb energy values that can exceed kinetic energy in magnetic field $\hbar\omega_c$; $\omega_c=eB/m^{\ast}$ and $m^{\ast}$ is effective electron mass.\cite{Goldman08} 

The possible importance of CB charging was suggested by earlier experiments of Ford \textit{et al.}\cite{Ford94} and Kataoka \textit{et al.},\cite{Kataoka00} where doubled frequency of the AB oscillations was observed. For the filling factor $f_c=2$, one may anticipate that resonances from one spin species should occur halfway between the neighboring resonances of the second spin species. This is, however, not the case. In a later experiment, using selective injection and detection of spin-resolved edge channels, it was shown that the antidot states with up spin do not provide resonant paths in the $h/2e$ AB oscillations.\cite{Kataoka03} No model of noninteracting electrons can explain this, because in such models the $h/2e$ oscillations should be a simple composition of the two $h/e$ oscillations coming from the two spin species, and the phase shift between the two $h/e$ oscillations is determined by the ratio between the Zeeman energy and the single-particle level spacing. Since the ratio depends on the antidot potential at the Fermi level and the magnetic field, the noninteracting model cannot provide an explanation of the sample-independent $\pi$ phase shift. Moreover, in the absence of interactions, both spins should participate in the resonant scattering, contradicting the experimental observation that only the spin species with the larger Zeeman energy contributes to the resonances.\cite{Ford94, Kataoka00, Kataoka03} Thus, experimental findings gave strong motivation for a model that takes electron interactions into account. 

To explain double-frequency Aharonov-Bohm oscillations, models accounting for the formation of compressible rings around the antidot\cite{Kataoka00} and capacitive interaction between excess charges were introduced\cite{Sim03}. The first model is based on the assumption that there are two compressible regions encircling an antidot, separated by an insulating incompressible ring. Screening in compressible regions, and Coulomb blockade, then force the resonances through the outer compressible region to occur twice per $h/e$ cycle.\cite{Kataoka00} In the capacitive-interaction model,\cite{Sim03} two antidot-bound edge states are assumed to localize excess charges that are spatially separated from each other and from extended edge channels by incompressible regions. This allows one to include in the antidot Hamiltonian the capacitive coupling of excess charges. In a regime of weak coupling, Coulomb blockade prohibits relaxation of the excess charges unless one of the antidot states accumulates exactly one electron or spin-flip cotuneling between them is allowed. Analyzing the evolution of the excess charges as a function of magnetic field, it was proposed that the process responsible for doubling of the AB oscillations comprises two consecutive tunneling events of spin-down electrons and one intermediate Kondo resonance.\cite{Sim03} This type of process agrees with the experiment\cite{Kataoka03} where only the electrons with spin down contribute to the $h/2e$ AB oscillations. 

A related topic of $1/f_c$ periodic AB oscillations has been recently investigated both experimentally\cite{Camino, Marcus09} and theoretically\cite{Halperin07, QDinterferometer} for the case of quantum dot-based interferometers. The experiments of Camino \textit{et al.}\cite{Camino} clearly demonstrated that the magnetic flux AB period is described by Eq. \eqref{period_f_c} and the gate voltage period stays constant for any filling factor $f_c$. Moreover, the authors reanalyzed the existing experimental data and showed that all of it can be also well described by Eq. \eqref{period_f_c}. To explain $1/f_c$ scaled period of AB oscillations in quantum dots, Coulomb blockade theory was introduced in Ref. \onlinecite{Halperin07}. Assuming that a compressible island exists inside the quantum dot, the AB period is caused by charging of $f_c$ fully occupied LLs in the dot. The validity of this assumption for the compressible island as well as the electrostatics of the AB interferometer has been discussed in Ref. \onlinecite{QDinterferometer} by two of the present authors. It was shown and explained why the scattering theory based on Landauer formula\cite{Davies_book} predicts the conventional AB periodicity, Eqs. \eqref{period_1}, \eqref{period_Vg}. A very recent experiment of Zhang \textit{et al.}\cite{Marcus09} unambitiously pointed out that $1/f_c$ period oscillations, that are caused by the CB effect, hold in an AB interferometer with a small quantum dot. However, as the dot size increases, the charging energy becomes an unimportant energy scale and the conventional AB oscillations are restored. 

In the present paper, motivated by experiment of Goldman \textit{et al.},\cite{Goldman08} we address electron transport through the antidot AB interferometer from different standpoints. Starting from a geometrical layout of the device, we calculate self-consistently the edge state structure, energy level evolution, charge quantization and linear-response conductance in the Thomas-Fermi and Hartree approximations. We find that the AB periodicity is well described by the conventional formulas \eqref{period_1}, \eqref{period_Vg} in the case of the ideal structure without impurities. The conductance is dictated by the highest occupied ($f_c+1$)-th state in the constrictions, but the $f_c$ antidot-bound states are well localized and do not participate in transport. Electron interactions in the Hartree approximation pin the $f_c$ single-particle states to the Fermi level and force their mutual positions to be correlated. The Hartree approach also predicts that the number of electrons around the antidot oscillates in a saw-tooth manner reflecting sequential escape of electrons from the $f_c$ edge states. For low temperatures, we have found an interesting phenomenon that we call anti-resonance reflection of the Fano type. While the Hartree and Thomas-Fermi approximations do not reproduce experimental $1/f_c$ AB periodicity,\cite{Goldman08} we explore two mechanisms that might be relevant, namely scattering by impurities and Coulomb blockade tunneling.

\section{Model}

\begin{figure*}[tb]
\includegraphics[keepaspectratio,width=\textwidth]{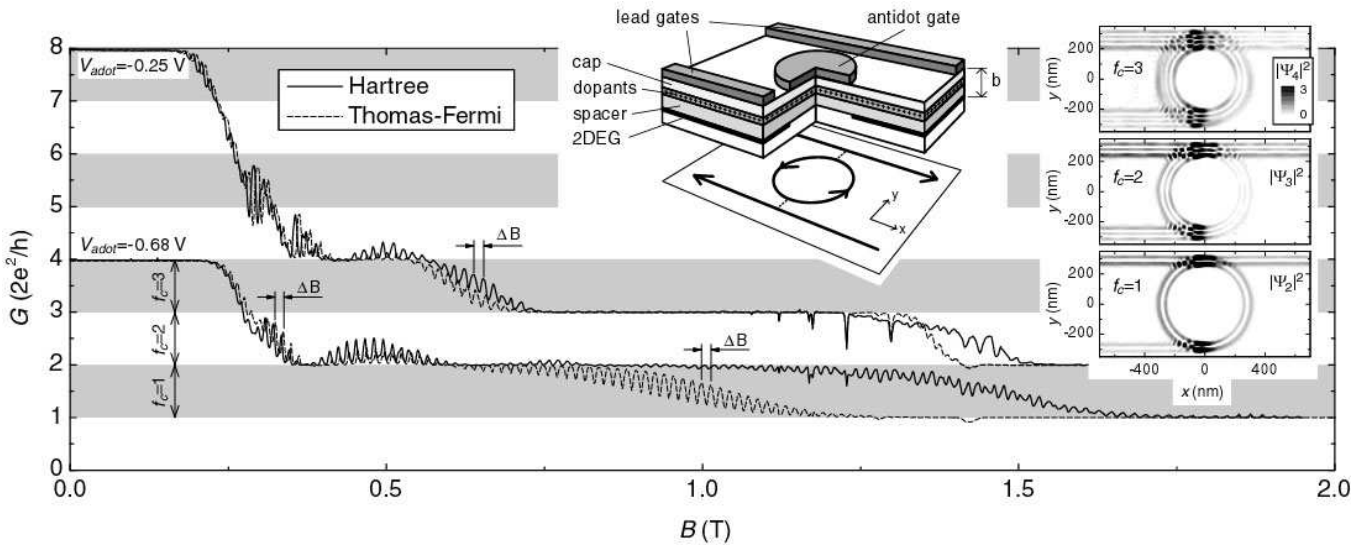}
\caption{The AB conductance oscillations calculated in the Hartree (solid lines) and Thomas-Fermi (dashed lines) approximations for two antidot gate voltages $V_{adot}=-0.25$, $-0.68$ V and for temperature $T=0.2$ K. $\Delta B$ marks the AB period, which is the same for any filling factor in the constriction $f_c$. Insets on the right show the wave function modulus for different $f_c$. Inset on the top shows schematic structure of the antidot-based AB interferometer. Top pattern denotes the metallic gates on the GaAs heterostructure. The radius of the antidot gate is $R=200$ nm. 2DEG resides a distance $b$ from the surface.} 
\label{fig:GvsB}
\end{figure*}

We consider an antidot AB interferometer defined by split-gates in the GaAs heterostructure similar to those studied experimentally.\cite{Sim_review, Hwang91, Ford94, Sachrajda94, Kataoka99, Goldman08} A schematic layout of the device is illustrated in Fig. \ref{fig:GvsB}. Charge carriers originating from a fully ionized donor layer form the two-dimensional electron gas (2DEG), which is buried inside a substrate at the GaAs/Al$_x$Ga$_{1-x}$As heterointerface situated at a distance $b$ from the surface. Metallic gates placed on the top define the antidot and the leads at the depth of the 2DEG.
 
The Hamiltonian of the whole system, including the semi-infinite leads, can be written in the form $H=H_{0}+V(\mathbf{r})$, where
\begin{equation}
	H_{0}=-\frac{\hbar ^{2}}{2m^{\ast }}\left\{ \left( \frac{\partial }{\partial
x}-\frac{eiBy}{\hbar }\right) ^{2}+\frac{\partial ^{2}}{\partial y^{2}}%
\right\}
\end{equation}%
is the kinetic energy in the Landau gauge, and the total confining potential
\begin{equation}
	V(\mathbf{r})=V_{conf}(\mathbf{r})+V_{H}(\mathbf{r}),
\end{equation}%
where $V_{conf}(\mathbf{r})$ is the electrostatic confinement (including contributions from the top gates, the donor layer and the Schottky barrier), $V_{H}(\mathbf{r})$ is the Hartree potential,
\begin{equation}
	V_{H}(\mathbf{r})=\frac{e^{2}}{4\pi \varepsilon _{0}\varepsilon _{r}}\int d%
\mathbf{r}\,^{\prime }n(\mathbf{r}^{\prime })\left( \frac{1}{|\mathbf{r}-%
\mathbf{r}^{\prime }|}-\frac{1}{\sqrt{|\mathbf{r}-\mathbf{r}^{\prime
}|^{2}+4b^{2}}}\right) ,  
	\label{V_H}
\end{equation}%
where $n(\mathbf{r})$ is the electron density, the second term corresponds to the mirror charges situated at the distance $b$ from the surface, $\varepsilon _{r}=12.9$ is the dielectric constant of GaAs. The antidot and the leads are treated on the same footing, i.e. the electron interaction and the magnetic field are included both in the lead and in the antidot regions.\cite{opendot} 

We calculate the self-consistent electron densities, potentials and the conductance on the basis of the Green's function technique. The description of the method can be found in Refs. \onlinecite{opendot, QPC, Zozoulenko_1996} and thus the main steps in the calculations are only briefly sketched here. First we compute the self-consistent solution for the electron density, effective potential and the Bloch states in the semi-infinite leads by the technique described in Ref. \onlinecite{wire}. Knowledge of the Bloch states allows us to find the surface Green's function of the semi-infinite leads. We then calculate the Green's function of the central section of the structure by adding slice by slice and
making use of the Dyson equation on each iteration step. Finally we apply the Dyson equation in order to couple the left and right leads with the central section and thus compute the full Green's function $\mathcal{G}(E)$ of the whole system. The electron density is integrated from the Green's function (in the real space),
\begin{equation}
n(\mathbf{r}) = -\frac{1}{\pi }\int_{-\infty }^{\infty }\Im \lbrack \mathcal{G}(\mathbf{r},\mathbf{r},E)] f_{FD}(E-E_{F})dE,
\end{equation}%
where $f_{FD}$ is the Fermi-Dirac distribution. This procedure is repeated many times until the self-consisten solution is reached; we use a convergence criterium $\left\vert n_{i}^{out} - n_{i}^{in} \right\vert / (n_{i}^{out} + n_{i}^{in}) < 10^{-5}$, where $n_{i}^{in}$ and $n_{i}^{out}$ are input and output densities on each iteration step $i$.

Finally the conductance is computed from the Landauer formula, which in the linear response regime is\cite{Davies_book}
\begin{equation}
	G = -\frac{2e^{2}}{h} \int_{-\infty }^{\infty }dET(E) \frac{\partial f_{FD}(E-E_{F})}{\partial E},
	\label{conductance}
\end{equation}
where the transmission coefficient $T(E)$, is calculated from the Green's function between the leads.\cite{opendot, QPC, Zozoulenko_1996} 

To clarify the role of the electron interaction we also calculate the conductance of the open dot in the Thomas-Fermi (TF) approximation where the self-consistent electron density and potential are given by the semiclassical TF equation at zero field
\begin{equation}
	\frac{\pi \hbar^2}{m^{\ast}}n(\mathbf{r})+V(\mathbf{r})=E_F.
	\label{TF}
\end{equation}
This approximation does not capture effects related to electron-electron interaction in quantizing magnetic field such as formation of compressible and incompressible strips and hence it corresponds to noninteracting one-electron approach where, however, the total confinement is given by a smooth fixed realistic potential.\cite{opendot, statistics}

While the present approach is not expected to account for single-electron tunneling in the conductance (leading to the Coulomb blockade peaks),\cite{Datta04} one can expect that it correctly reproduces a global electrostatics of the interferometer and microscopic structure of the quantum mechanical edge states regardless whether the conductance is dominated by a single-electron charging or not. This is because the interferometer is an open system with a large number of electrons surrounding the antidot and thus the electrostatic charging caused by a single electron hardly affects the total confining potential of the interferometer. Thus the results of the self-consistent Hartree approach provide accurate information concerning the locations of the propagating states and the structure of compressible/incompressible strips in the interferometer. Our calculations are also expected to provide detailed information concerning the coupling strengths between the states in the leads and around the antidot.  

\section{Results and discussion}

We calculate the magnetotransport of a quantum antidot AB interferometer with following parameters representative of a typical experimental structure.\cite{Sim_review, Hwang91, Ford94, Sachrajda94, Kataoka99, Goldman08} The 2DEG is buried at $b=50$ nm below the surface (the widths of the cap, donor and spacer layers are 14 nm, 26 nm and 10 nm respectively), the donor concentration is $1.02\times 10^{24}$ m$^{-3}$. The width of the quantum wire and semi-infinite leads is 700 nm. The radius of the antidot gate is $R=200$ nm, see Fig. \ref{fig:GvsB}. The gate voltage applied to the lead gates is $V_{lead}=-0.4$ V. With these parameters of the device there are 25 channels available for propagation in the leads and the electron density in the center of the leads is $n_{lead}=2.5\times 10^{15}$ m$^{-2}$. 

\subsection{Magnetic flux periodicity of Aharonov-Bohm oscillations}

Figure \ref{fig:GvsB} shows the conductance of the AB interferometer as a function of magnetic field calculated for the quantum-mechanical Hartree and semi-classical Thomas-Fermi approximations. At very low magnetic field the conductances of two parallel QPC constrictions are additive and behave like Ohmic resistors.\cite{George90} At high magnetic field, in the quantum Hall regime, the total conductance exhibits step-like dependence caused by gradual depopulation of the Landau levels (LLs).\cite{Davies_book} All channels for electron propagation are fully open or fully blocked in the plateau regions, while a partly transmitted channel is present in the transition regions. The conductance oscillations due to Aharonov-Bohm interference are clearly seen in the transition regions, which indicates that they are caused by the ($f_c+1$)-th partly transmitted channel. 

The Aharonov-Bohm oscillations reveal the same period $\Delta B = 15$ mT independent of $f_c$ for both the Hartree and Thomas-Fermi approximations as is seen in Fig. \ref{fig:GvsB}. This periodicity is in excellent agreement with the conventional AB formula \eqref{period_1}. It corresponds to an area enclosed by skipping orbits of 0.28 $\mu$m, which gives a circle radius of $r=300$ nm. The latter is slightly larger that the geometrical radius of the antidot gate $R=200$ nm and is in agreement with the wave function maps in the insets in Fig. \ref{fig:GvsB}. 

\begin{figure}[tb!]
\includegraphics[keepaspectratio,width=\columnwidth]{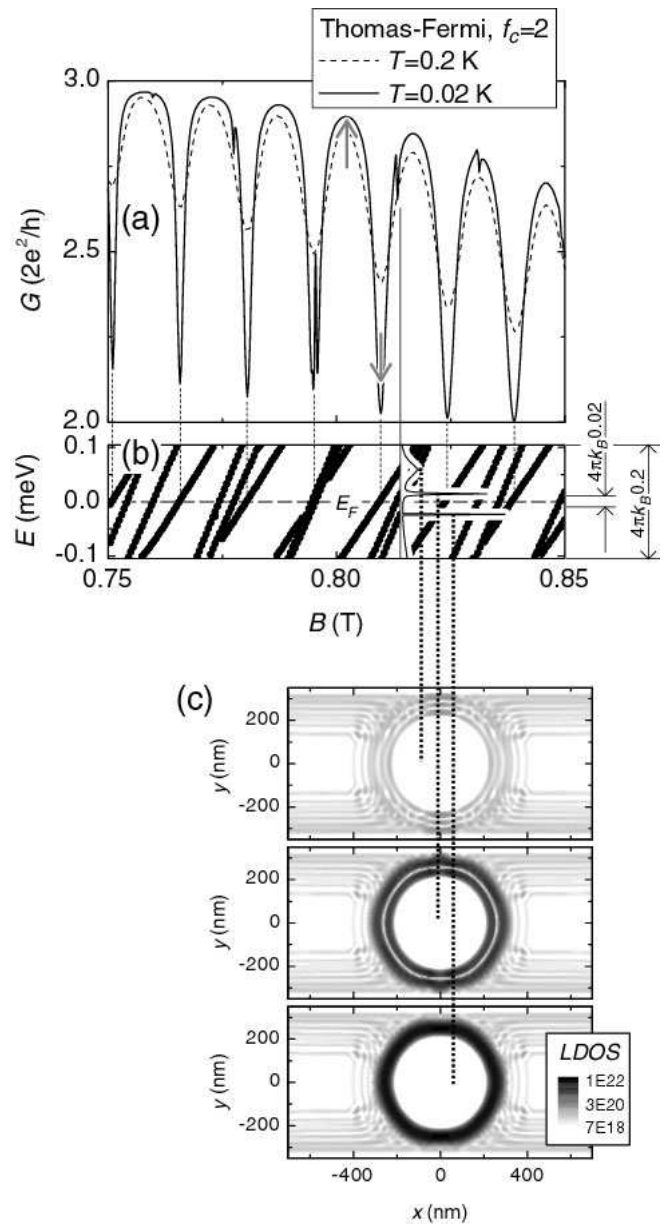}
\caption{(a) The conductance calculated in the Thomas-Fermi approximation for different temperatures and $V_{adot}=-0.4$ V; solid curve: 0.02 K, dashed curve: 0.2 K. (b) Evolution of the resonant energy levels near $E_F$. The inset shows $DOS$ in a ring around the antidot for the specified value of magnetic field $B=0.814$ T. The ring is chosen so that contains the antidot-bound states and has outer and inner radii equal to 350 nm and 200 nm, respectively (note that the antidot gate radius is $R=200$ nm). The evolution of the energy levels was obtained from the peak positions of the $DOS$ at each given value of $B$. (c) $LDOS$ at three peaks corresponding to three different energies at $B=0.814$ T.} \label{fig:resonances}
\end{figure}

The AB oscillations can be related to evolution of the corresponding energy spectrum when single-electron
states cross the Fermi level each time the flux, through the antidot, increases by the flux quantum. This is illustrated in Fig. \ref{fig:resonances}(b), which shows an evolution of the resonant levels as a function of magnetic field in the
vicinity of the Fermi energy. [To obtain the evolution of the resonant levels we draw an imaginary ring and analyze the density of states (DOS) there at each given $B$. The inner radius of the ring is chosen to be the antidot radius and the outer radius is 150 nm larger, i.e. we account for all states in the range $200...350$ nm from the center. When DOS has been calculated, its peaks are searched for and their positions are plotted as illustrated in Fig. \ref{fig:resonances}(b)]. Each conductance minimum of the AB interferometer seen in Fig. \ref{fig:resonances}(a) corresponds to a resonant level aligns with the Fermi energy $E_F$ in \ref{fig:resonances}(b), implying a condition of resonants reflection of the extended edge state in the lead by the antidot.\cite{Jain88} However, two out of every three states present in Fig. \ref{fig:resonances}(b) are associated with neither minimum nor maximum of the AB conductance oscillations at 0.2 K (the dashed curve in Fig. {fig:resonances}(a)). Only if the temperature is lowered from 0.2 K to 0.02 K, does the conductance start to reveal a feature caused by their presence at $E_F$. Inspection of the $DOS$ at $B=0.814$ T, see the inset in Fig. \ref{fig:resonances}(b), shows that these extra states produce very narrow and sharp resonances (with broadening $\Gamma << k_B T$ for $T=0.2$ K), while the resonance due to the conductance-mediating state is broad and low ($\Gamma \sim k_B T$). Let's call two former states ''antidot-bound states`` and refer to the latter one as a ''transport state``. The single-particle transport state originates from the partly transmitted ($f_c+1$)-th LL in the constriction. It is strongly coupled to the extended edge states in the leads. By contrast, all $f_c$ antidot-bound states are weakly coupled to the extended edge states, because they are situated closer to the antidot hill and surrounded by an incompressible strip. Figure \ref{fig:resonances}(c) displays $LDOS$ maps clearly showing the presence of $f_c=1$ and $f_c=2$ antidot-bound states at corresponding resonance peaks in $DOS$. Note that evolution of both the antidot-bound and transport states is not correlated between each other. All of them cross the Fermi energy at different magnetic fields although the $\Delta B$ interval for single-particle states belonging the same LL stays constant. This is a characteristic feature of noninteracting electrons in the Thomas-Fermi approximation. 

\subsection{Anti-resonance reflection}

\begin{figure}[tb!]
\includegraphics[keepaspectratio,width=\columnwidth]{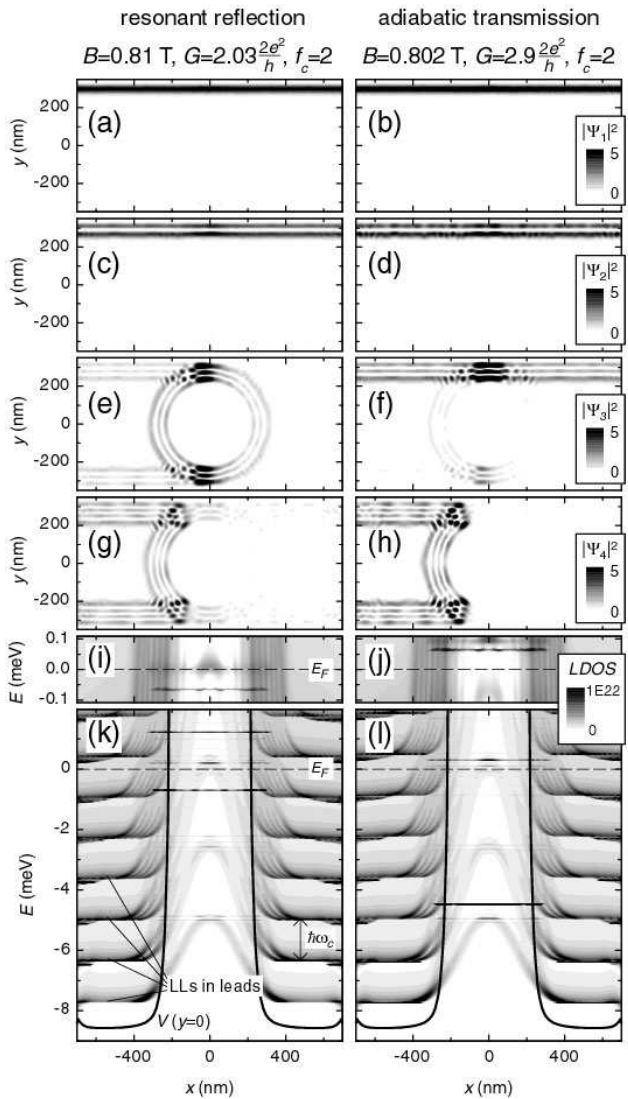}
\caption{The AB resonant reflection and adiabatic transmission at $f_c=2$ as calculated in the Thomas-Fermi approximation and marked by arrows in Fig. \ref{fig:resonances}(a). The left column corresponds to minimum and right column to maximum of the conductance, $B=0.81$ T and 0.802 T, respectively. Plots (a)-(h) show the wave function modulus $\left| \Psi_i \right|^2$ of $i$-th edge state and (i)-(l) are for $LDOS$ integrated over transverse $y$-direction. Panels (i) and (j) illustrate resonant levels in the vicinity of $E_F$ in an enlarged scale of (k) and (l). Fat solid lines in (k)-(l) are the total confinement potential along $y=0$. } 
\label{fig:basics}
\end{figure}

To understand features that are seen to be superimposed on the AB conductance oscillations in Fig. \ref{fig:resonances}(a), let's first look at the wave functions and $LDOS$ at minima and maxima of the conventional AB oscillations due to the ($f_c+1$)-th transport state. These are shown in Fig. \ref{fig:basics}: the wave functions for $f_c$ lead edge states show perfect transmission through the device, while the state $f_c+2$ as well as all higher states exhibit perfect reflection of the incoming state from the left lead back into the left lead. The only state modulating the transport through the antidot is the transport state $f_c+1$. It effectively provides a resonant tunneling pathway between incoming and outgoing edge channels via the antidot-bound state. Inspection of $LDOS$ in Figs. \ref{fig:basics}(i)-(l) shows presence of the single-particle state at the Fermi energy during resonant reflection via ($f_c+1$)-th transport state.

\begin{figure}[tb!]
\includegraphics[keepaspectratio,width=\columnwidth5]{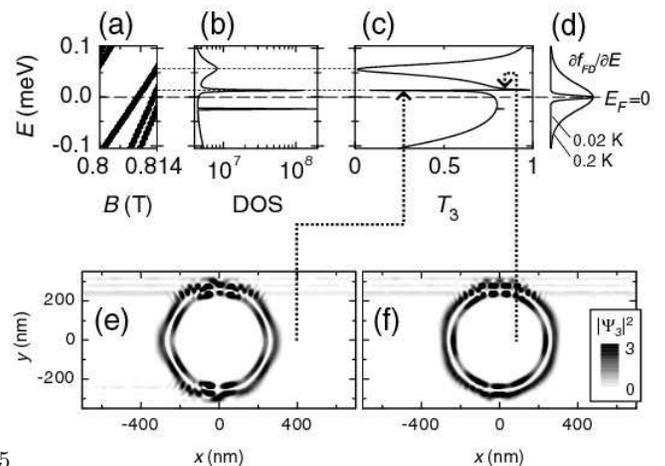}
\caption{(a),(b) Fragment of energy structure and $DOS$ at $B=0.814$ T from Fig. \ref{fig:resonances}(b). (c) Transmission coefficient $T_3=\sum_i T_{i3}$ from third mode in the left lead to all available $i$-th modes in the right lead. There is a reflection anti-resonance at $E=0.014$ meV that manifests itself as a zigzag jump in $T_3$. (d) Derivative of the Fermi-Dirac function for $T=0.02$ and 0.2 K. (e),(f) Third wave function squared modulus, $\left| \Psi_3 \right|^2$, at two sides of the anti-resonance.} 
\label{fig:antiresonance}
\end{figure}

At low temperatures, the $f_c$ antidot-bound states produce a sharp zigzag-like feature in the conductance as shown in Fig. \ref{fig:resonances}(a) at $T=0.02$ K. This effect is most pronounced for $B=0.79$ T and might be attributed to a Fano-type resonance.\cite{Fano61} It differs qualitatively from the common resonance dip due to ($f_c+1$)-th transport states. In order to get insight into its origin, let's look at the transmission coefficient $T_{f_c+1}$ vs. energy plot, Fig. \ref{fig:antiresonance}(c). Note that the conductance is an integral over the total transmission weighted by the Fermi-Dirac derivative, Eq. \eqref{conductance}, and thus it can not resolve all of the details unless the temperature is extremely low. Overall the dependence of $T_{f_c+1}$ is quite smooth showing a deep minimum associated with resonant reflection of ($f_c+1$)-th state. This state can also be monitored in $DOS$ as the wide and low peak, Fig. \ref{fig:antiresonance}(b). However, there is a superimposed zigzag-like jump in $T_{f_c+1}$, which is totally different and caused by scattering between the $f_c+1$)-th state and the $f_c$ one. It happens at $E=-0.014$ meV in Fig. \ref{fig:antiresonance}(c) and is caused by $3 \leftrightarrow 2$ scattering. At an energy slightly less than $E=-0.014$ meV, there is a sharp dip, which is associated with the third extended state being well coupled to the second antidot-bound state, Fig. \ref{fig:antiresonance}(e). However, as energy increases and slightly exceeds $E=-0.014$ meV, it turns into a sharp peak with the antidot-bounded state becoming perfectly isolated, Figs. \ref{fig:antiresonance}(f). The fact that it isolated can be seen as absence of ``bridges'' to the extended lead state and almost perfect circled shape. Because the shake reveals in the transmission, this feature is not indent in the conductance at temperatures exceeding $k_B T > \Gamma$. We refer this type of resonance to as an anti-resonance of the Fano-type\cite{Fano61} since it results from quantum interference between two processes, one involving strongly localized state, the $f_c$ antidot orbital. When the incident energy exactly coincides with anti-resonance energy, the AB phase $2\pi \phi / \phi_0$ flips by $\phi=\pi$. It worth mentioning that the antiresonance here differs from phase change in the AB conductance oscillations observed in Refs. \onlinecite{Ford94, Sachrajda94, George94}, where a phase flip between consecutive oscillations accompanied by change of oscillation period.\cite{Sachrajda94, George94}

\subsection{Effect of electron interaction in Hartree approximation}

\begin{figure}[tb!]
\includegraphics[keepaspectratio,width=\columnwidth]{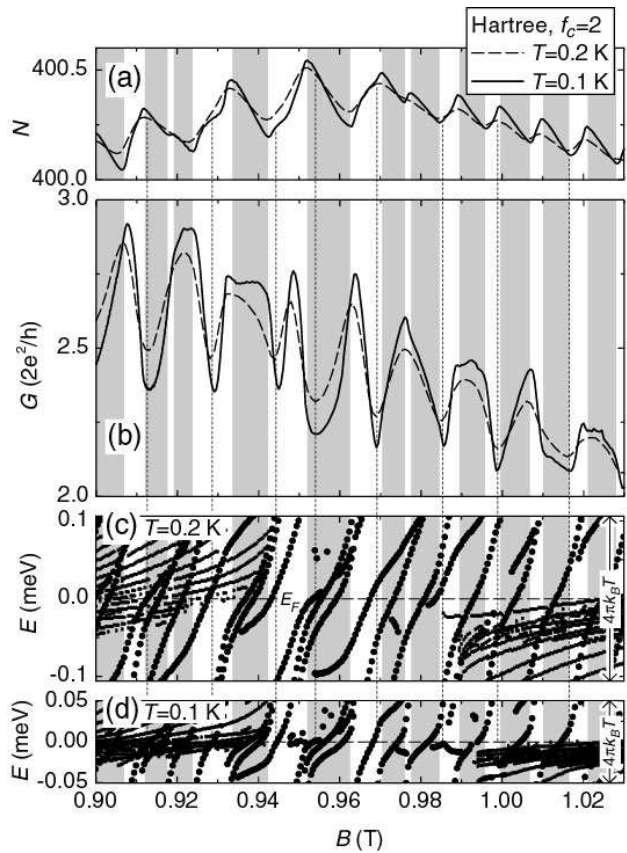}
\caption{Number of electrons around quantum antidot $N$ (a), the AB conductance oscillations (b) and evolution of the resonant energy levels near $E_F$ (c),(d) calculated in the Hartree approximation for different temperatures and $V_{adot}=-0.4$ V. $N$ and energy levels are calculated in an annulus around the antidot whose outer and inner radii equal to 350 nm and 200 nm, respectively (note that the antidot gate radius is $R=200$ nm). Evolution of the energy levels in (c) is calculated at $T=0.2$ K, while (d) is for $T=0.1$ K. The multiple levels marked by smaller circles are due to bulk states in the leads, which inevitably captured in the region of interest, i.e. the annulus around the antidot.} 
\label{fig:Hartree}
\end{figure}

Accounting for electron interactions within the quantum-mechanical Hartree approximation brings qualitatively new features to the AB oscillations, Fig. \ref{fig:Hartree}. First, the conductance doesn't show perfect smooth regular oscillations any longer: their shapes become very distorted, which is especially pronounced as temperature lowers. Secondly, the energy levels from both $f_c+1$ and $f_c$ edge states become correlated: their positions at $E_F$ become more equally spaced. This is attributed to the effect of Coulomb interaction that favors only one single-particle state being depopulated at a given magnetic field. In other words, electrons escape from localized $f_c$ states one by one. Third, as the temperature decreases the energy levels due to $f_c$ edge states become pinned to $E_F$, Figs. \ref{fig:Hartree}(c),(d). We define the pinning as a lower slope at $E_F$; see Ref. \onlinecite{opendot} for detailed discussion of the pinning effect. When a state is pinned to $E_F$, it easily adjusts its position or occupation in response to any external perturbation. The role of the perturbation might be played by either magnetic field or applied gate voltage. Therefore, the screening of the antidot, and related metallic-like behavior of the system, is provided solely by the $f_c$ states, not by the ($f_c+1$)-th transport state. The latter crosses $E_F$ steeply and mediates the conductance oscillations in a similar fashion to that in the semi-classical Thomas-Fermi approximation. Fourth, the number of electrons $N$ in an annulus around the antidot shows saw-tooth oscillations that reflects pinning and depopulation of $f_c$ edge states. Intervals of magnetic field with linear negative slopes of $N$ and pinned states are marked by shaded regions in Fig. \ref{fig:Hartree}. The negative slopes of $N$ in Fig. \ref{fig:Hartree}(a) are caused by gradual depopulation when the corresponding single-particle state is pushed up in energy and its occupation decreases. Note that the change of electron number is less than unity, which we explain by the finite temperature and bulk states captured in the region of interest, i.e. in the annulus of $200...350$ nm size from the center. It also worth noting that the saw-tooth dependence of $N$ is in excellent agreement with the experimental findings in Ref. \onlinecite{Kataoka99}. 

While the shape of the AB conductance oscillations is strongly non-sinusoidal in the Hartree approximation, their periodicity still described by the conventional Eq. \eqref{period_1} that, in turn, disagrees with the experiment of Goldman \textit{et al}.\cite{Goldman08} The Hartree approach is known to describe well the electrostatics of the system at hand. This is confirmed by the good agreement with numerous experiments including, for example, formation of compressible/incompressible strips in quantum wires\cite{wires} and the statistics of conductance oscillations in open quantum dots.\cite{statistics} Thus, the validity of the energy level evolution as well as the electron number oscillation presented in Fig. \ref{fig:Hartree} is indeed qualitatively correct. The conductance, however, may be incorrect. It is calculated using the Landauer-Buttiker formalism that has been shown to fail in the regime of weak coupling, when the conductance is less than the conductance quantum $G_0 = 2e^2/h$.\cite{Datta04} Though the total conductance is larger than $G_0$ the antidot is in the weak coupling regime (a related discussion of quantum dot-based AB interferometers is given in Ref. \onlinecite{QDinterferometer}). This is because of the adiabatic character of the transport when the lowest $f_{c}$ states pass thought the interferometer without any reflection, see Fig. \ref{fig:basics}. The highest ($f_{c}+1$)-th transport edge state, giving rise to the AB oscillations in the transition regions between the plateaus, becomes thus effectively decoupled from the remaining $f_{c}$ states. Therefore, because of well localized $f_{c}$ states and partly localized ($f_{c}+1$)-th state, the electron charge in the antidot may become quantized and transport through the interferometer strongly affected by the Coulomb blockade effect. 



\subsection{Gate voltage periodicity of Aharonov-Bohm oscillations}

\begin{figure}[tb!]
\includegraphics[keepaspectratio,width=\columnwidth]{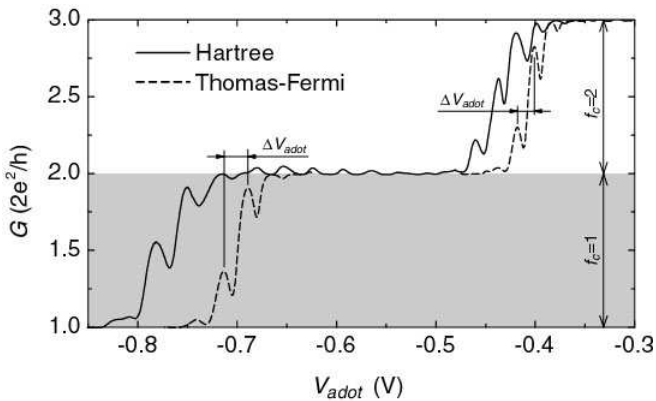}
\caption{The AB conductance oscillations calculated in the Hartree (solid lines) and Thomas-Fermi (dashed lines) approximations for fixed magnetic field $B=0.8$ T. $\Delta V_{adot}$ is the AB period, which scales linearly with the filling factor in the constriction $f_c$. Temperature $T=0.2$ K.} 
\label{fig:GvsV}
\end{figure}

The Aharonov-Bohm oscillations can be also observed when the gate voltage varies for fixed magnetic field. Figure \ref{fig:GvsV} shows the conductance as a function of the antidot gate voltage calculated in the Thomas-Fermi and Hartree approximations for $B=0.8$ T. As in the case of magnetic field dependence, Fig. \ref{fig:GvsB}, the AB conductance oscillations are pronounced in the transition regions between plateaus. However, the period $\Delta V_{adot}$ scales linearly with the filling factor $f_c$, in agreement with the conventional formula \eqref{period_Vg}. 

Both magnetic field and gate voltage periodicity of the AB oscillations are well described by conventional formulas \eqref{period_1}, \eqref{period_Vg}. However, it disagrees with the experiment, Ref. \onlinecite{Goldman08}. To overcome this discrepancy, we consider two effects in the following, namely the scattering by random impurity potentials and the Coulomb blockade theory.

\subsection{Scattering by random impurity potentials}

It is known that impurity scattering might substantially modify electron transport in AlGaAs heterostructures.\cite{Davies_book} Depending on the nature of scatters, the scattering potential varies widely. Charged impurities such as ionized donors have a long-range potential, whereas neutral impurities have short-range potentials. These two cases have different effects on electron transport. The first leads to localization of edge states in the quantum Hall regime and strong modification of the transition regions between quantum Hall plateaus.\cite{impQHE} The edge states circulating around the antidot might change their locations and, as a result, the AB oscillations might experience sudden period changes.\cite{Sachrajda94, George94} Because $1/f_c$ periodicity in the experiment of Goldman \textit{et al.}\cite{Goldman08} is robust and measured for different cooldown cycles, we are skeptical that the long-range scattering is responsible for $1/f_c$ periodicity and concentrate in the following on the short-range scattering. 

\begin{figure}[tb!]
\includegraphics[scale=0.45]{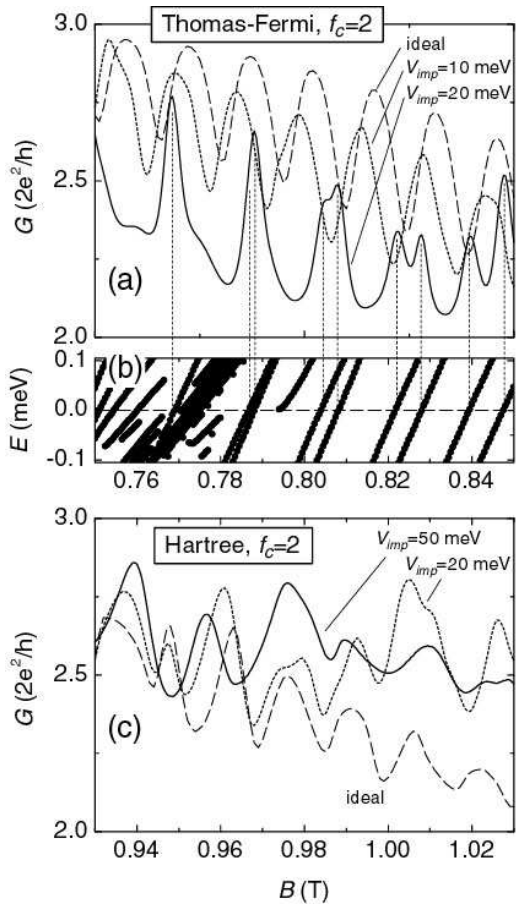}
\caption{The conductance (a) and resonant energy structure (b) calculated in the Thomas-Fermi approximation for various disorder potentials $V_{imp}$. (c) The conductance calculated in the Hartree approximation. $V_{adot}=-0.4$ V. }
\label{fig:imp}
\end{figure}

A physical realization of the short-range disorder potential occurs if some neutral impurity such as Al atom has diffused out of a AlGaAs barrier into a GaAs well where the 2DEG resides.  The Al atoms are the scattering centers with a potential $V_{imp}$. Because we are interesting in scattering between the extended and bound states and between different bound and partly-bound states, we generated random point scatters in the vicinity of the antidot. Figures \ref{fig:imp}(a),(b) show the conductance and energy level structure calculated in the Thomas-Fermi approximation for the case of 500 random scatters, which corresponds to concentration $n_{imp}=10^{15}$ m$^{-2}$. As the magnitude of $V_{imp}$ increases, the scattering between different edge states becomes stronger. The AB oscillations due to resonant reflection of the ($f_c+1$)-th state are gradually suppressed with a new oscillation pattern emerging from resonant transmission via the $f_c$ antidot-bound states. The positions of the AB conductance peaks are clearly correlated with the antidot-bound states crossing $E_F$, see Figs \ref{fig:imp}(a),(b). The transport ($f_c+1$)-th state, which governs the conductance in the ideal case without impurities, becomes easily destroyed because it is half-filled in the QPC constrictions and, thus, slight potential fluctuations effectively block its propagation. However, in contrast to the ideal case with no defect scattering, resonant transmission peak is visible in Fig. \ref{fig:imp}(a) when ever any antidot bound state crosses $E_F$ in Fig. \ref{fig:imp}(b) for strong defect scattering at higher values of the magnetic field that are shown.

The Hartree approximation, Fig. \ref{fig:imp}(c), does not clearly recover resonant transmission triggered by the disorder potential, as it occurs in the Thomas-Fermi approach. In this case, there are only faint remnants of resonant transmission via the $f_c$ antidot-bound states at $V_{imp}=20$ meV. We attribute this to inadequate modeling of point defects due to Al atoms. The minimal area that can be occupied by a defect in our simulations is $5\times5$ nm$^2$. This is two orders of magnitude larger than the realistic cross section of the Al atom. On the other hand, the height of the realistic defect potential is also much larger and of the order of $V_{imp}^{Al}\approx1$ eV. Note that the screened potential after self-consistent calculation is about ten times smaller than the input $V_{imp}$ in Fig. \ref{fig:imp}(c). If we perform simulation for higher potentials the self-consistent density becomes quickly washed out preventing electron transport through the region occupied by the defects. Thus, we conclude that the short-range disorder is a plausible source of $1/f_c$ AB periodicity, but a quantitative comparison remains to be done and is beyond the scope of the present work. 

\subsection{Coulomb blockade model for Aharanov-Bohm oscillations}

The quantum antidot does not confine electrons electrostatically, but sufficiently large magnetic field causes formation of localized bound states where charge might be quantized. Direct evidence of the charging effect in the antidot was given in the experiment of Kataoka \textsl{et al.}\cite{Kataoka99} Placing a noninvasive voltage probe in close proximity to the antidot, they detected steady accumulation followed by sudden relaxation of a localized excess charge nearby. The saw-tooth resistance oscillations measured by the detector coincide with the resonances monitored in the antidot conductance. Therefore, it was concluded that a source of the excess charge is the antidot, and its conductance is mediated by Coulomb charging. Additional evidence for the CB effect follows from measurement of the conductance as functions of both magnetic field and source-drain bias, where clear and regular Coulomb diamonds were observed.\cite{Kataoka99} The doubled frequency conductance oscillations measured in Refs. \onlinecite{Ford94, Kataoka00} are also a strong indication of the Coulomb charging in the antidot. Thus, we conclude that $1/f_c$ periodicity observed by Goldman \textit{et al.}\cite{Goldman08} might be a result of the CB effect. As indirect support, it worth mentioning that rough estimation of the electron Coulomb interaction energy gives values exceeding kinetic energy in magnetic field.\cite{Goldman08} 

\begin{figure}[tb]
\includegraphics[keepaspectratio,width=\columnwidth]{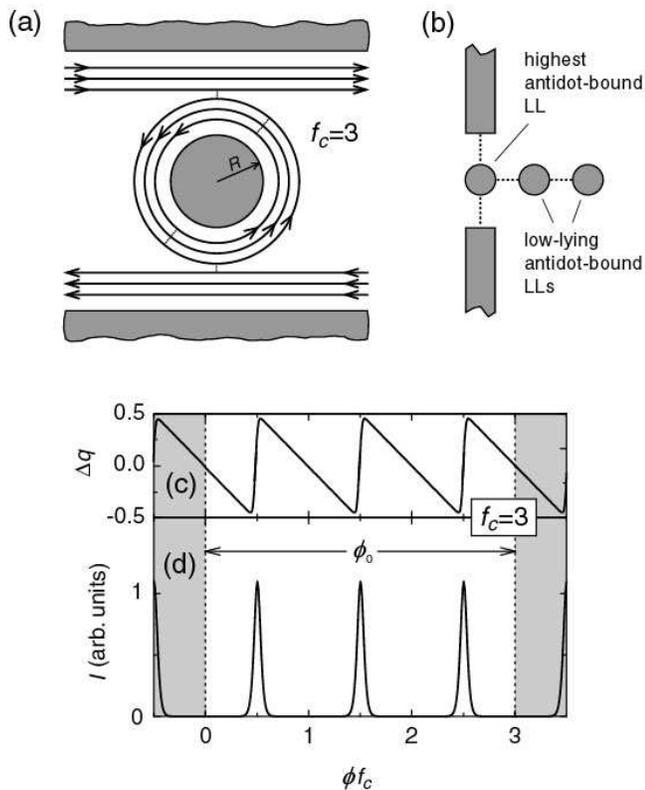}
\caption{(a) Schematic illustration of the edge states in a quantum antidot when the filling factor in constrictions is $f_c=3$. An electron from extended edge states may tunnel into the outermost antidot-bound state, but not into the bound states due to the low-lying LLs. (b) Equivalent single-electron scheme for the antidot edge states. Dotted lines mark paths for sequential tunneling. (c) Excess charge on the LLs, $\Delta q$, and (d) Coulomb blockade oscillations calculated within the orthodox theory.}
\label{fig:CB}
\end{figure}

Motivated by these experimental arguments as well as our calculation results presented above, we develop a simple phenomenological model based on the CB orthodox theory.\cite{Likharev99} Let's consider a case of $f_c$ fully occupied LLs in the QPC constrictions, when the conductance is near a plateau region. Notice that this case is represented in the experiment of Goldman \textit{et al.}\cite{Goldman08} Figure \ref{fig:CB}(a) illustrates schematically the edge states existing around the antidot. For a given magnetic field and antidot gate voltage we may draw a closed curve of area $S$ that encompasses all $f_c$ bound states. There is some background number of electrons $N$ inside the area $S$. If we fix the position of the curve and increase the field by amount $\Delta B = \phi_0 / S$, the total flux through the reference area, $\phi$, will increase by one flux quantum $\phi_0$. One single-particle state in each $f_c$ LL gets pushed up in energy, crosses the Fermi energy and becomes depopulated. The number of electrons in the reference area drops by $f_c$. However, it is known that the magnetic field does not change the number of electrons\cite{Davies_book} and, thus, electrons cannot just disappear. There must be a balancing influx of electrons into LLs in a way that the degeneracy of each LL increases exactly by one. For the case of a voltage applied to the antidot gate, a background charge in the reference area can be increased and $N_{gate}$ additional electrons can be attracted to the area $S$. Hence, we obtain a total charge imbalance inside the area $S$ given by $eN + frac\left(ef_c\phi\right) - eN_{gate}$, which leads to a charging energy 
\begin{equation}
	E=\frac{e^2}{2C}\left( N + frac\left(ef_c\phi\right) - N_{gate} \right)^2,
	\label{eq:energy}
\end{equation}
where $C$ is the capacitance of the island and $frac(x)$ means the fractional part of $x$. Note that the charging energy \eqref{eq:energy} takes the same value when $\phi$ changes by $1/f_c$. Therefore, magnetic field periodicity satisfies Eq. \eqref{period_f_c} as measured in the experiment of Goldman \textit{et al.}\cite{Goldman08} The antidot gate period, however, is one electron charge in the reference area independent of the filling factor. It is also worth noting a similar approach to CB charging for the quantum dot-based interferometer, Ref. \onlinecite{Halperin07}.

Several comments about the CB theory proposed above. First, the change of magnetic field or gate voltage is supposed not to be large so that the reference area $S$ always encloses a fixed number $f_c$ of edge states. Secondly, all edge states encircle the antidot at about the same radii from the center and the extent of their wave functions is an unimportant length scale. This is evidently correct for large antidot radii or large magnetic fields. Third, the charging energy \eqref{eq:energy} does not depend on which particular edge state builds the charge imbalance at a given field and gate voltage. Eq. \eqref{eq:energy} rather treats all $f_c$ states as one single-electron island. However, our calculations within both the Thomas-Fermi and Hartree approaches identify the highest LL as the most important for electron transport. On the other hand, experimental data clearly indicates that transport occurs via the highest LL (the outermost edge state) when the antidot in the CB regime.\cite{Kataoka03} Thus, we assume that a single-electron island implied in Eq. \eqref{eq:energy} has an internal structure and functions as a single-electron trap,\cite{Likharev99} see Fig. \eqref{fig:CB}(b). Electrons sequentially hop from the low-lying LLs into the extended edge states and vice versa via highest ($f_c$)-th LL. This is supported by electrostatics shown in Fig. \ref{fig:Hartree}(a), where states depopulate one by one. 


To solve the electron transport problem in the Coulomb blockade regime as governed by Eq. \eqref{eq:energy}, we employ the standard orthodox theory.\cite{Likharev99} It describes an evolution of system via a ``master'' equation for $p(N)$, the probability that there are $N$ electrons in the island
\begin{eqnarray}
	\label{eq:master}
	\Gamma_{N-1 \rightarrow N} \: p(N-1) + \Gamma_{N+1 \rightarrow N} \: p(N+1) = \\ \nonumber
	\left[ \Gamma_{N \rightarrow N-1} + \Gamma_{N \rightarrow N+1} \right] p(N).
\end{eqnarray}
Here $\Gamma_{N^{\prime} \rightarrow N}$ is the sum of the transition rates through tunnels barriers, which change the electron number $N^{\prime}$ to $N$. Each transition rate treats tunneling of a single electron through a tunnel barrier as a random event and depends on the reduction of the electrostatic energy of the system, Eq. \eqref{eq:energy}, resulting from such a tunneling event. For example, the transition rate from the source electrode to island reads
\begin{equation}
	\Gamma^{s\rightarrow i}(N) = \frac{\Delta E^{s\rightarrow i}(N)}{e^2 R_s} \left( 1 - e^{\Delta E^{s\rightarrow i}(N) / k_B T} \right)^{-1},
	\label{eq:gamma}
\end{equation}
where $\Delta E^{s\rightarrow i}(N)$ is the energy change after tunneling from the source to the island and $R_s$ is the resistance of the tunneling barrier between the source and island. We solve equation \eqref{eq:master} and then calculate the average current as 
\begin{equation}
	I = \sum_N \left( \Gamma^{s\rightarrow i}(N) - \Gamma^{i\rightarrow s}(N) \right) \: p(N).
\end{equation} 

Figures \ref{fig:CB}(c),(d) show a representative calculation for the case of $f_c=3$. We use capacitances and resistances of the tunnel barriers estimated from our self-consistent calculations given above. For a realistic quantum antidot shown in Fig. \ref{fig:GvsB}, approximate parameters are $C_1=C_2\approx10^{-18}$ F and $R_1=R_2\approx125$ k$\Omega$. The excess charge $\Delta q$ as a function of magnetic flux is shown in Fig. \ref{fig:CB}(c). It oscillates in a saw-tooth manner within a window $-0.5 < \Delta q < 0.5$. Each time the value $+0.5$ is approached, an electron tunnel through the potential barriers with no energy cost and substantial current starts flowing through the antidot. When the magnetic flux increases further, one of the LLs gets  recharged by one electron and the process repeats for an other LL. For $f_c=3$, increase of magnetic field by the flux quantum $\phi_0$ generates three successive rises of $\Delta q$ and three related peaks of the current. Note that the current is proportional to the resistance measured in a two-terminal setup. 

\section{Conclusions}

In the present paper, we provide a microscopic physical description of the edge states existing in the quantum antidot focusing on the related conductance oscillations due to the Aharonov-Bohm interference. Motivated by recent experiment of Goldman \textit{et al.},\cite{Goldman08} we discuss different mechanisms that might be a source of measured $1/f_c$ periodicity of the AB conductance oscillations. Our finding are summarized as follows.

1) Approaches based of the Hartree and Thomas-Fermi models for an ideal antidot structure predict the conventional AB conductance oscillations, as described by formulas \eqref{period_1}, \eqref{period_Vg}, i.e. magnetic field period does not depend on $f_c$ and gate voltage period scales linearly with $f_c$. This is caused by transport isolation of the $f_c$ edge states circulating around the antidot and the conductance being modulated solely by the highest occupied ($f_c+1$)-th state in the constrictions.

2) Electron interactions in the quantum-mechanical Hartree approximation bring qualitatively new features to the electrostatics of the antidot AB interferometer. The single-particle states originating from the $f_c$ edge states become pinned to the Fermi energy. Their mutual positions at the Fermi energy are correlated, such that there is only one single-particle state depopulating at a given magnetic field and gate voltage. The number of electrons around the antidot shows a related saw-tooth dependence. It reflects the fact that particles escape sequentially, i.e. one-by-one.

3) As the temperature decreases, both the Hartree and Thomas-Fermi approximations start to reveal a reflection anti-resonance of the Fano type. This manifests itself as zigzag jumps in the conductance, with the ($f_c+1$)-th transport state scattered into $f_c$ antidot-bound state and vice versa. On the lower energy side of the anti-resonance these states are strongly coupled with each other, but they are perfectly isolated on the high energy side.

4) The experimentally measured $1/f_c$ periodicity might be recovered if some disorder is introduced around the antidot. This can be naturally realized due to Al atoms diffused into the well where the 2DEG is. A short-ranged potential forces different edge states to mix and all $f_c$ states might start to participate in transport. This is accompanied by changing of the AB conductance from resonant reflection to resonant transmission. 

5) A simple Coulomb blockade theory might also explain the $1/f_c$ periodicity. Using information about edge states and their occupancy from the self-consistent calculations, we write down the charging energy of the system and calculate the transport in the sequential tunneling regime. When the magnetic field changes by a flux quantum, the system experiences recharging by $f_c$ electrons. 

While the experiment\cite{Goldman08} as well as our present study show presumably that $1/f_c$ AB periodicity is caused by CB tunneling, we suggest an experiment to verity that. If the size of the antidot were increased many times, the capacitance of the system would increase proportionally. The charging energy, therefore, would decrease and CB tunneling become unimportant.\cite{Marcus09} In this situation, the conventional AB oscillations, independent of $f_c$ if magnetic field changes, should be restored.

From the theoretical point of view, it would be interesting to calculate the conductance by the Hartree-Fock approach, where the self-interaction problem is eliminated. 
One might also try density functional theory with a proper exchange-correlation functional that avoids self-interaction errors as well. 

\begin{acknowledgments}
This work was supported by NSERC of Canada, the Canadian Institute for Advanced Research and the Swedish Research Council. Numerical calculations were performed using the facilities of the National Supercomputer Center, Link\"{o}ping, Sweden. 
\end{acknowledgments}

\end{document}